\documentclass{birkart04}
 \theoremstyle{definition}
 
 \theoremstyle{remark}

 \numberwithin{equation}{section}

\newcommand{\vr}{\vec{r}}
\input epsf
\begin{document}
%
%
%
%
\title[]{Configurational Continuum modelling of crystalline surface evolution}
\author{Navot Israeli}
\address{Department of Physics of Complex Systems, Weizmann Institute of Science, Rehovot, 76100, Israel.}

\email{navot.israeli@weizmann.ac.il}

\author{Daniel Kandel}
\address{Department of Physics of Complex Systems, Weizmann Institute of Science, Rehovot, 76100, Israel.}
\email{daniel.kandel@weizmann.ac.il}



\date{January 1, 2004}

\begin{abstract}
We propose a novel approach to continuum modelling of dynamics of
crystal surfaces. Our model follows the evolution of an ensemble
of step configurations, which are consistent with the macroscopic
surface profile. Contrary to the usual approach where the
continuum limit is achieved when typical surface features consist
of many steps, our continuum limit is approached when the number
of step configurations of the ensemble is very large. The model is
capable of handling singular surface structures such as corners
and facets and has a clear computational advantage over discrete
models.
\end{abstract}

\maketitle
\section{Introduction}
The behavior of classical physical systems is typically described
in terms of equations of motion for discrete microscopic objects
(e.g.\ atoms). The dynamics of the microscopic objects is usually
very erratic and complex. Nevertheless, in many cases a smooth
behavior emerges when the system is observed on macroscopic length
and time scales (e.g.\ in fluid flow through a pipe). A
fundamental problem in physics is to understand the emergence of
the smooth macroscopic behavior of a system starting from its
microscopic description. A useful way to address this problem is
to construct a continuum, coarse-grained model, which treats the
dynamics of the macroscopic, smoothly varying, degrees of freedom
rather than the microscopic ones. The derivation of continuum
models from the microscopic dynamics is far from trivial. In most
cases it is done in a phenomenological manner by introducing
various uncontrolled approximations.

In this work we address the above problem in the context of the
dynamics of crystal surfaces. The evolution of crystal surfaces
below the roughening transition proceeds by the motion of discrete
atomic steps which are separated by high symmetry orientation
terraces. One can model step motion by solving the diffusion
problem of adatoms on the terraces with appropriate boundary
conditions at step edges. This approach was introduced long ago by
Burton, Cabrera and Frank \cite{BCF}, and was further developed by
other authors \cite{WilliamsReviews}. The resulting models specify
the normal velocity of each step in the system as a function of
its position and shape and as a function of position and shapes of
neighboring steps. These step flow models are capable of
describing surface evolution on the mesoscopic scale with
significant success \cite{Fu_PRL77,Tanaka_PRL78}. However, step flow
models pose a serious challenge for numerical computations, and
can be solved only for small systems.

Several attempts were made to construct continuum models for
stepped surfaces
\cite{Mullins_JAP28,Ozdemir_PRB42,Nozieres_JPI48,Lancon_PRL64,Uwaha_JPSJ57,Chame_BulChemComm,HagerSpohn_SurfSci324,cone_prl,cone,sine_scaling,1D_scaling,Bonzel_AppPhys35,Bonzel_SurfSci336,Ramana_PRB62},
in order to understand their large scale properties. The general
idea behind these attempts is that step flow can be treated
continuously in regions where every morphological surface feature
is composed of many steps. If we label surface steps by the index
$n$, the continuum limit in these models is obtained by taking $n$
to be continuous. In what follows we will refer to these models as
the {\it conventional} approach.

In the literature, there are two methods to derive conventional
continuum models. One method is to write down the discrete step
equations of motion and then transform them into a partial
differential equation by taking the step index $n$ to be
continuous
\cite{Nozieres_JPI48,cone_prl,cone,sine_scaling,1D_scaling}. The
second method is to start with a continuous surface free energy
density and derive a surface dynamic equation that minimizes
it\cite{Ozdemir_PRB42,Lancon_PRL64,HagerSpohn_SurfSci324,Bonzel_AppPhys35,Bonzel_SurfSci336,Ramana_PRB62}.
These two methods are complementary provided that: 1. The free
energy density of the second method is the continuum analog of the
free energy of an array of discrete steps. 2. The two methods uses
the same mass transport mechanism. Such continuum models are
fairly successful in describing the evolution of smooth surfaces
with very simple morphologies. However, they suffer from
fundamental drawbacks, which do not allow generalizations to more
complex and realistic situations.

The most severe drawback is that below the roughening temperature,
crystal surfaces have singularities in the form of corners and
macroscopic facets. The latter are a manifestation of the cusp
singularity of the surface free energy at high symmetry crystal
orientations. The assumption that every surface feature is
composed of many steps clearly breaks down on macroscopic facets
where there are no steps at all. Thus, existing continuum models
fail conceptually near singular regions. Several authors have
tried to overcome this problem by solving a continuum model only
in the non-singular parts of the surface and then carefully match
the boundary conditions at the singular points or
lines\cite{HagerSpohn_SurfSci324,sine_scaling,cone,1D_scaling}. In most cases
however it is not at all clear how these matching conditions can
be derived. This difficulty is a fundamental drawback of
conventional continuum models and not merely a technicality. As we
argue below, boundary conditions at the singular points or lines
cannot be derived in the context of conventional continuum models.

To see why this is true consider the situation near a facet edge.
The step at the facet edge is special and obeys a unique equation
of motion. In contrast to steps in the sloping parts of the
surface which all have two neighboring steps of the same sign, a
facet step has only one neighbor of the same sign and potentially
a second neighbor of an opposite sign. There might also be special
physical conditions such as surface reconstruction that add to the
uniqueness of a facet step. As we found in several cases
\cite{cone_prl,cone,sine_scaling,1D_scaling}, the unique behavior
of a facet step sensitively determines the amount of material
emitted or absorbed at the facet and the rate at which steps cross
the facet and annihilate. When going to the continuum limit these
quantities serve as flux boundary conditions at the singularity.
The problem is that conventional continuum models are derived from
the equations of motion (or from the surface free energy density)
that apply away from the facet and are therefore ignorant of the
special behavior of facet steps. Thus, the boundary conditions at
the facet edge must be derived from a careful analysis of the
discrete dynamics of faces steps. However, in going to the
continuum in the conventional way, one loses the information
regarding the position of individual steps and the discrete
analysis cannot be performed.

Another approach for dealing with surface singularities is to
round the surface free energy cusp
\cite{Bonzel_AppPhys35,Bonzel_SurfSci336,Ramana_PRB62},
approximating true facets by relatively flat but analytic regions.
This method avoids the need of specifying explicit boundary
conditions at the singularity by assuming analyticity of the
surface. The correct surface behavior is then expected to be
captured in the limit of vanishing cusp rounding. This procedure
completely ignores the key role of facet steps and implicitly
assumes that the surface free energy derived for non singular
orientations determines the dynamics on facets as well. This
assumption is generally false due to the same reasons discussed
above. An example for a case where cusp rounding leads to
erroneous results can be found in Appendix
\ref{cusp_rounding_appendix}.

In this work we propose a conceptually new definition of the
continuum limit, which we term Configurational
Continuum\cite{prl_configcont}. Configurational Continuum allows
construction of continuum models, which are free of all the
limitations of conventional continuum models discussed above. It
provides a rigorous way of deriving the continuum model directly
from the discrete step equations of motion. Like other continuum
models, Configurational Continuum has a clear computational
advantage over the discrete step model due to the small number of
discretization points it requires for the description of smooth
surface regions in a numerical scheme.

\section{Configurational continuum}
\label{config_cont_introduction} 
\begin{figure}[h]
\centerline{ \epsfxsize=80mm \epsffile{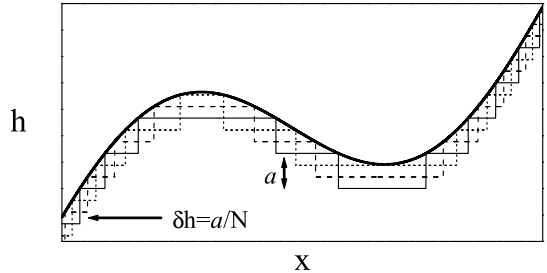}} \caption{A
schematic illustration of the ensemble of configurations whose
upper envelope defines the continuum limit.} \label{ensemble}
\end{figure}

In order to overcome the
limitations of conventional continuum models we now propose a
conceptually new definition of the continuum limit for stepped
surfaces. Our key observation is that a continuous surface height
profile can be represented by many similar, but not identical,
step configurations. Figure \ref{ensemble} is a one dimensional
demonstration of this point. It shows a continuous height
function, $h(x)$, of position $x$ (thick solid line), and three
valid microscopic representations of this profile as step
configurations. The main idea of this work is to define the height
profile in the continuum limit as the upper envelope of the
discrete height functions of an {\em ensemble} of many such step
configurations.

To construct the configurations of the ensemble let $a$ be the
height of a single step and $N$ the number of configurations in
the ensemble. We construct the ensemble so that the height
difference $\delta h$, between two adjacent configurations is
$a/N$, as depicted in Fig.\ \ref{ensemble}. The continuum limit is
obtained when $N \rightarrow \infty$. The generalization to higher
dimensions is straightforward.

The dynamics of the continuum model is as follows. Each step
configuration of the ensemble evolves according to the microscopic
dynamics. As a result, the envelope of discrete height functions
changes with time, thus defining the evolution of the continuous
height function $h(\vr)$ where $\vr$ is a vector in the high
symmetry $xy$ plane. There is one technical complication which
might arise if two steps from different configurations cross each
other. Such an event would make $h(\vr)$ a multi valued function
of position and requires a more general mathematical description
of the surface. For simplicity we ignore this and assume that
$h(\vr)$ remains single valued.

There is a crucial assumption hidden in this definition of the
continuum model. We postulate that our construction leads to a
mathematically well defined height function at all times. When
does this assumptions hold? Consider two initially similar
configurations of the ensemble. Our continuum limit is well
defined provided these two configurations have similar microscopic
dynamics and hence remain similar at later times. Note however
that this assumption has to hold in the conventional continuum
definition as well, and therefore does not put additional
restrictions on our model. In fact, if two initially similar
configurations evolve very differently, one must abandon the
continuum limit and follow the specific microscopic configuration
of interest, using discrete dynamics.

We now derive the evolution equation for the continuous height,
$h(\vr,t)$, at position $\vr$ and time $t$. As a basis for the
derivation we assume knowledge of the discrete equations of motion
for the underlying step flow model. These equations of motion
specify the normal velocity of a step that passes through
$(\vr,t)$ as a function of the local step configuration. We denote
this dependency by writing
\begin{equation}
\vec{v}\left(\vr,t\right)=\vec{v}\left(C_{\vr,t}\right),
\label{general_step_flow}
\end{equation}
where $C_{\vr,t}$ denotes the configuration of steps in the region
that influences the step velocity at $(\vr,t)$. In most models of
step flow this region of influence covers a small number of
neighboring steps. Note that in the context of the discrete step
model, $C_{\vr,t}$ is the actual configuration of steps in the
system. When going to the continuum we will be interested in the
ensemble of configurations $\{C_{\vr,t}\}$ which are consistent
with $h(\vr,t)$.

The continuous height $h(\vr,t)$ changes with time due to the flow
of steps through $\vr$, and due to nucleation and annihilation of
steps. At this stage we disregard nucleation processes and include
them later. First, we consider positions which are not local
extrema of the height profile. It is obvious from the construction
of the Configurational Continuum, that for each point $\vr$ there
is exactly one configuration $C_{\vr,t}$ in the ensemble, which
has a step that passes through $\vr$ at time $t$. That step lies
along the unique equal-height contour line, which passes through
$\vr$. As is demonstrated in Fig.\ \ref{ensemble}, the exact
positions of neighboring steps in the configuration $C_{\vr,t}$
can be calculated from the knowledge of $h(\vr,t)$, and the fact
that in this configuration there is a step at $\vr$. Hence, we can
use the discrete step model Eq.\ (\ref{general_step_flow}) and
calculate the normal velocity of the step
$\vec{v}\left(C_{\vr,t}\right)$. Note that at different positions,
$\vec{v}\left(C_{\vr,t}\right)$ is the normal velocity of steps
which may belong to {\em different} configurations in the
ensemble.

Next we define the directional gradient in the direction from
which steps flow towards $\vr$
\begin{equation}
\nabla h_{-\hat{v}}(\vr,t)\equiv -\hat{v}(\vr,t)\lim_{\epsilon\to
0_+}\frac{h\left(\vr-\epsilon
\cdot\hat{v}(\vr,t)\right)-h\left(\vr,t\right)}{\epsilon}\;,
\end{equation}
where
$\hat{v}(\vr,t)=\frac{\vec{v}\left(C_{\vr,t}\right)}{\left|\vec{v}\left(C_{\vr,t}\right)\right|}$.
This is useful for the calculation of the current of steps
arriving at $\vr$:
\begin{equation}
J(\vr,t)=\frac{N}{a}\left|\nabla h_{-\hat{v}}(\vr,t)\cdot
\vec{v}\left(C_{\vr,t}\right)\right|\;,
\end{equation}
where we have used the fact that the local step density is
$\left|\nabla h_{-\hat{v}}(\vr,t)\right|N/a$. Note that $J$ is the
current of steps belonging to all configurations in the ensemble,
and not to one particular configuration. Since each step (from any
configuration), which passes through $\vr$ changes the height of
the ensemble envelope by $a/N$, the continuous height profile
obeys the evolution equation
\begin{equation}
\frac{\partial h(\vr,t)}{\partial t}=-\nabla
h_{-\hat{v}}(\vr,t)\cdot \vec{v}\left(C_{\vr,t}\right)\;.
\label{dynamic_equation}
\end{equation}

The above derivation of the evolution in the continuum is not
valid at local extrema of the surface, because generally one
cannot define a unique equal-height contour line which passes
through such a point. To avoid the problem, we define $\partial
h/\partial t$ at local extrema as the limit of the height time
derivative as one approaches these points. This limiting procedure
is justified, since there are no microscopic realizations of the
surface with steps exactly at the local extrema.

At this point we emphasize that the Configurational Continuum
evolution is formally identical to the evolution of the discrete
step model. This statement is almost trivial, since the definition
of Eq.\ (\ref{dynamic_equation}) follows the envelope of the
ensemble of configurations, and each configuration evolves with
step velocities calculated from the discrete step model. Thus Eq.\
(\ref{dynamic_equation}) is exact. Moreover, the use of
directional derivatives in the derivation of Eq.\
(\ref{dynamic_equation}) makes it valid even at singular surface
regions such as corners or facet edges. Similarly to other
continuum models it is solved numerically by discretization of
space, which is the only approximation involved in such solutions.


What is the relation between Configurational Continuum and
conventional continuum models? In regions where $h(\vr,t)$ is
analytic, the evolution equation (\ref{dynamic_equation}) reduces
to the continuity equation
\begin{equation}
\frac{\partial h(\vr,t)}{\partial t}=- \nabla h(\vr,t) \cdot
\vec{v}\left(C_{\vr,t}\right)\;. \label{analytic_dyn_eq}
\end{equation}
If $h(\vr,t)$ is sufficiently smooth,
$\vec{v}\left(C_{\vr,t}\right)$ can be approximated as a local
function of $h$ and its spatial derivatives, as is commonly done
in conventional continuum models. Making this approximation will
therefore recover the conventional continuum approach. We can
conclude that in analytic surface regions the conventional
continuum approach approximates the Configurational Continuum
model and that the approximation quality depends on the smoothness
of the surface. However, near corners, facets or regions where the
profile is not smooth, one cannot reconstruct the microscopic step
configuration from the local value of $h$ and its spatial
derivatives. In these regions $\vec{v}\left(C_{\vr,t}\right)$
contains non local information and as a result Eq.\
(\ref{dynamic_equation}) cannot even be written as a differential
equation.

Is there any computational gain in using such a continuum model?
After all, we replaced a discrete model, which follows the
evolution of a single microscopic step configuration, by a model
which follows a whole ensemble of step configurations. The key
point is that we do not have to follow all the steps of all
configurations. To calculate $\partial h(\vr,t)/ \partial t$, it
is enough to locally follow the single configuration which has a
step that passes trough $\vr$ at time $t$. In addition, the
continuum evolution equation is solved on a grid, and the density
of grid points can be very small in regions where the continuous
height profile is smooth. The smoothness of the profile allows
very accurate interpolation between these points. Only near
singular points or lines we have to use a rather dense grid, and
there is no computational gain in these regions. In practice, the
total number of grid points used can be orders of magnitude
smaller than the number of points one has to use in order to
follow the evolution of a single microscopic step configuration.

So far we ignored the possibility of island or void nucleation. It
is possible to include island or void nucleation in our model
provided that we have a microscopic description for these events
which determines the nucleation probability in a given step
configuration. Within our continuum approach, the nucleation
probability at a point on the continuous surface is the ensemble
average of the microscopic nucleation probabilities at this point.
For demonstration proposes we consider a simple model where the
probability for the nucleation of an island on a terrace grows as
the square of the local concentration of diffusing adatoms.
Information regarding the values of terrace adatom concentrations
is already contained in the underlying step flow model Eq.\
(\ref{general_step_flow}), since it is used in the calculation of
adatom fluxes into and out of steps.

\section{Numerical solutions of the Configurational Continuum}
\label{config_cont_application} We now apply our approach to a few
simple cases. First, we consider a conic structure which consists
of circular concentric steps. This crystalline cone was studied in
Refs.\ \cite{cone_prl,cone} where we wrote a one-dimensional step
flow model for the step radii in the absence of growth flux. In
the diffusion limited case where adatom diffusion across terraces
is the rate limiting process, the equation of motion for the step
radii $r_n$ read:
\begin{eqnarray}
\frac{dr_1}{dt}&=&\frac{D_s C^{eq}_{\infty} \Omega}{k_B T
r_1}\frac{\mu_1-\mu_2}{\ln\left(r_1/r_2\right)}\;,
\nonumber \\
\frac{dr_n}{dt}&=&\frac{D_s C^{eq}_{\infty} \Omega}{k_B T
r_n}\left(\frac{\mu_n-\mu_{n+1}}{\ln\left(r_n/r_{n+1}\right)}-\frac{\mu_{n-1}-\mu_{n}}{\ln\left(r_{n-1}/r_n\right)}\right)\;,
\;\;\; n>1\;, \label{cone_step_velocity}
\end{eqnarray}
with the step chemical potentials $\mu_n$ given by
\begin{equation}
\mu_n = \frac{\Omega \Gamma}{r_n} + \Omega G \left[
  \frac{2r_{n+1}}{r_{n+1}+r_n} \cdot
  \frac{1}{\left(r_{n+1}-r_n\right)^3} -
  \frac{2r_{n-1}}{r_n+r_{n-1}} \cdot
  \frac{1}{\left(r_n-r_{n-1}\right)^3}
\right]\;. \label{cone_step_chemical_potential}
\end{equation}
In the above expressions $D_s$ is the adatom diffusion constant,
$\Omega$ is the atomic area of the crystal and $C^{eq}_{\infty}$
is the equilibrium concentration in the vicinity of a straight
isolated step. $\Gamma$ is the step line tension, $G$ is the
step-step interaction strength, $T$ is the temperature and $k_B$
is the Boltzmann constant. Eq.\
(\ref{cone_step_chemical_potential}) can be used for calculating
the chemical potential of the top step, $\mu_1$, by omitting the
second interaction term.

Numerical integration of the above step model shows that the cone
decays through the collapse of individual islands. During the
decay a facet develops at the top of the cone with a radius that
grows with time as $t^{1/4}$. Similar equations can be written in
the presence of growth flux. With flux, the cone grows except at
the peak which initially decays and then saturates. A facet forms
at the peak after saturation.

The continuum model we solved in the context of this example is a
fully two dimensional model, which can, in principle, develop non
radially symmetric morphologies. The microscopic dynamics we used
were a two dimensional generalization of the microscopic equations
for step radii of the discrete one dimensional model given above.
In particular the step chemical potential was generalized to
\begin{equation}
\mu \left(\vr \right)=\Omega \Gamma \kappa\left(\vr\right)+\Omega
G \left( \frac{\exp \left[ \frac{ \left| \vr_d-\vr \right| \cdot
\kappa \left( \vr_d \right) }{2} \right] }{ \left| \vr_d-\vr
\right|^3}-\frac{\exp \left[ -\frac{ \left| \vr_u-\vr \right|
\cdot \kappa \left( \vr_u \right) }{2} \right] }{ \left| \vr_u-\vr
\right|^3} \right)\;, \label{exp_chemical_potential}
\end{equation}
where $\kappa\left(\vr\right)$ is the local step curvature.
$\vr_d$ and $\vr_u$ are the coordinates at which the lower and
upper neighboring steps  are closest to the step at $\vr$. It is
assumed that these neighbors have the same sign as the step at
$\vr$ and that steps of opposite signs do not interact (in which
case the relevant interaction terms in Eq.\
(\ref{exp_chemical_potential}) are omitted). This expression was
derived assuming an $l^{-2}$ repulsion between two segments of two
different steps which are separated by a distance $l$. Under this
assumption, Eq.\ (\ref{exp_chemical_potential}) is exact to first
order in the curvature of neighboring steps and qualitatively
captures the interaction when they have large curvatures.

The equations describing adatom diffusion on terraces were solved
assuming the steps at $\vr$, $\vr_d$ and $\vr_u$ are circular and
concentric and that the radius of the step at $\vr$ is $1/
\kappa(\vr)$. Any microscopic dynamics, such as a full solution of
the diffusion equation on each terrace, or a more detailed
calculation of the interactions between steps can easily be used
in the framework of our model. For the sake of demonstrating the
validity of Configurational Continuum the simple dynamics we chose
are sufficient.

\begin{figure}[h]
\centerline{ \epsfxsize=80mm \epsffile{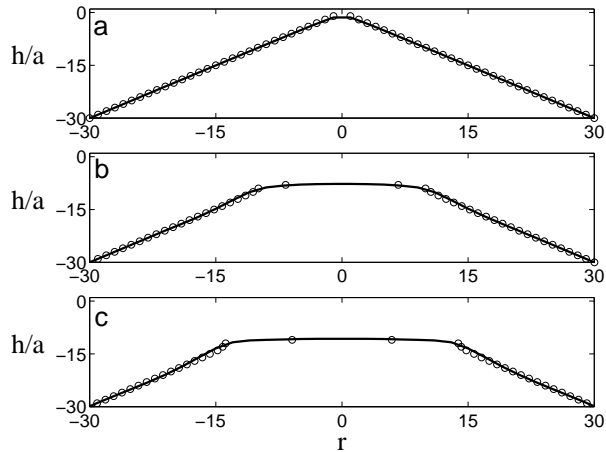}}
\caption{Decay of a crystalline cone. The solid line shows a cross
section of the two dimensional solution of Eq.\
(\ref{dynamic_equation}). Circles show the surface evolution
according to the one dimensional step flow model Eq.\
(\ref{cone_step_velocity}). (a) is the initial morphology and (b)
and (c) show the surface at later times.} \label{cone_decay}
\end{figure}

\begin{figure}[h]
\centerline{ \epsfxsize=80mm \epsffile{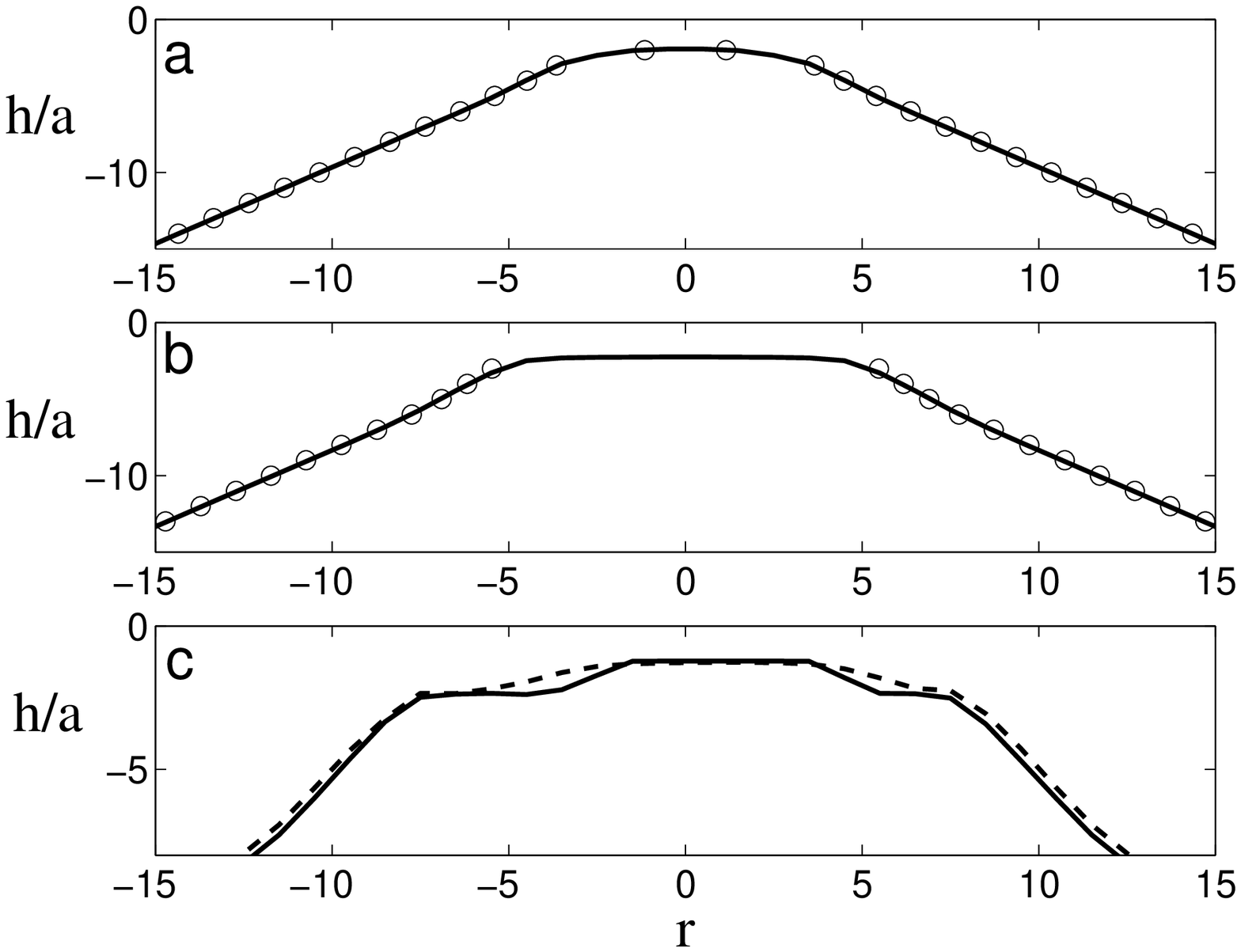}}
\vspace{1mm} \caption{Evolution of a crystalline cone under growth
conditions. The lines show cross sections of the two dimensional
solution of Eq.\ (\ref{dynamic_equation}). Circles show the
surface evolution according to the one dimensional step flow
model. The initial shape is the exact cone shown in Fig.\
\ref{cone_decay} (a). (a) and (b) show the generation and
evolution of a facet at the top at later times. (c) shows an
island which nucleated (solid) and grew (dashed) on the top facet.
Here no comparison is made with the step flow model.}
\label{growing_cone}
\end{figure}

Figure \ref{cone_decay} shows a comparison between a numerical
solution of the discrete step model and a cross section from the
two dimensional solution of our continuum model in the absence of
growth flux and when nucleation of new steps is not allowed.
Clearly the continuum model captures the main features of the
surface evolution. In particular the width and height of the top
facet are in excellent agreement with the discrete model. Fig.
\ref{growing_cone} shows a similar comparison in the presence of
growth flux. Again the agreement is quite impressive. In this case
we allowed new islands to nucleate and our simulation indicates
that nucleation events occur on the top facet once it becomes
large enough. There is hardly any nucleation on the finite slope
regions, because the steps there efficiently absorb the excess
material. Figure \ref{growing_cone}(c) shows an island on the top
facet, which nucleated and started growing.

We now turn to the more demanding example of bidirectional
sinusoidal grating relaxation. Here the initial surface height
profile of wave length $L$ is given by
\[
h_L(\vec{r},t=0)=h_0 \sin\left(\frac{2\pi x}{L}\right)
\sin\left(\frac{2 \pi y}{L}\right)\;.
\]
The relaxation of this profile towards a flat surface was studied
by Rettori and Villain \cite{Rettori_Vilain}, who gave an
approximate solution to a step flow model, in the limit where the
interaction between steps can be neglected with respect to the
step line tension. The decay of bidirectional sinusoidal profiles
was also studied numerically
 \cite{Ramana_PRB62}. We now apply our model to this problem assuming diffusion
limited kinetics without deposition flux or island nucleation.

For weak interactions between steps, the surface height evolves
according to $\partial h_L/\partial t\propto \nabla^2 \kappa \sim
L^{-3}$, where $\kappa$ is the step curvature. We therefore expect
the following scaling law for $h_L(\vec{r},t)$:
\begin{equation}
h_L(\vec{r},t)=h_{L=1}(\vec{r}/L,t/L^3)\;. \label{shape_scaling}
\end{equation}

\begin{figure}[h]
\centerline{ \epsfxsize=80mm \epsffile{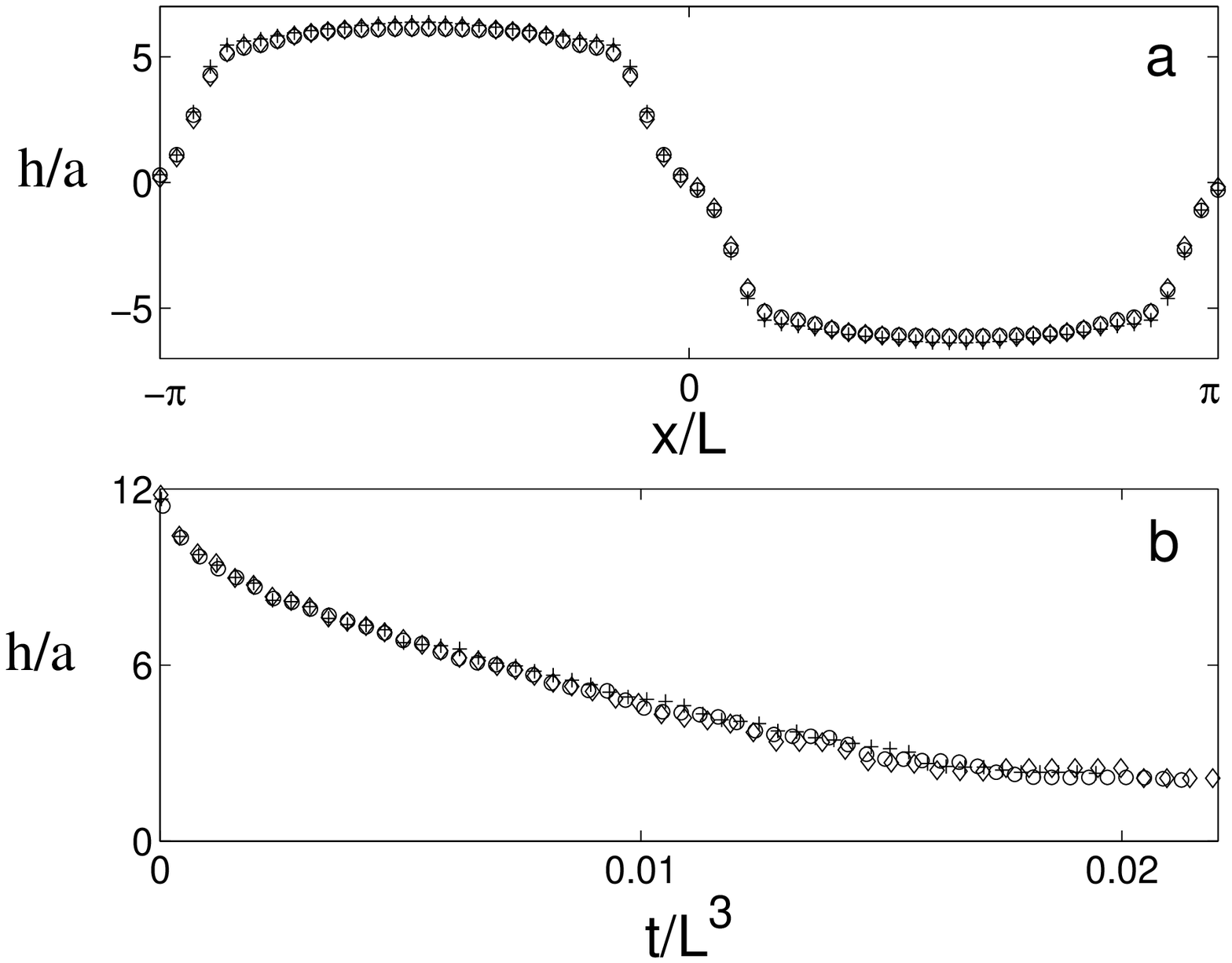}}
\vspace{1mm} \caption{Data collapse in the evolution of
bidirectional sinusoidal profiles with different wave lengths when
the repulsive interaction between steps is very weak. Wave lengths
shown are $L=64$ (circles), $L=128$ (squares) and $L=256$
(triangles). a) Cross sections of the profiles with different wave
lengths measured at time $t=t_0 L^3$. The cross sections are along
the $y=-L/4$ line (peak to valley). b) Amplitude decay of the
different wave lengths as a function of scaled time $t/L^3$.}
 \label{low_g_egg}
\end{figure}

Figure \ref{low_g_egg} shows the data collapse of cross sections
of profiles resulting from our continuum model. The different
symbols correspond to profiles of different wave lengths at time
$t=t_0 L^3$ for some fixed $t_0$. The quality of the data collapse
shows that the scaling scenario (\ref{shape_scaling}) holds very
accurately. Note that large facets have developed at the surface
extrema, and they are connected by very steep slopes. This shape
does not agree with Rettori and Villain's heuristic argument
\cite{Rettori_Vilain}, which predicts that facets appear also near
$h=0$ lines. Nevertheless, their prediction that after a short
transient the amplitude of the height profile decays as $t/L^3$ is
in agreement with both Eq.\ (\ref{shape_scaling}) and our
numerical solutions.

\begin{figure}[h]
\centerline{ \epsfxsize=80mm \epsffile{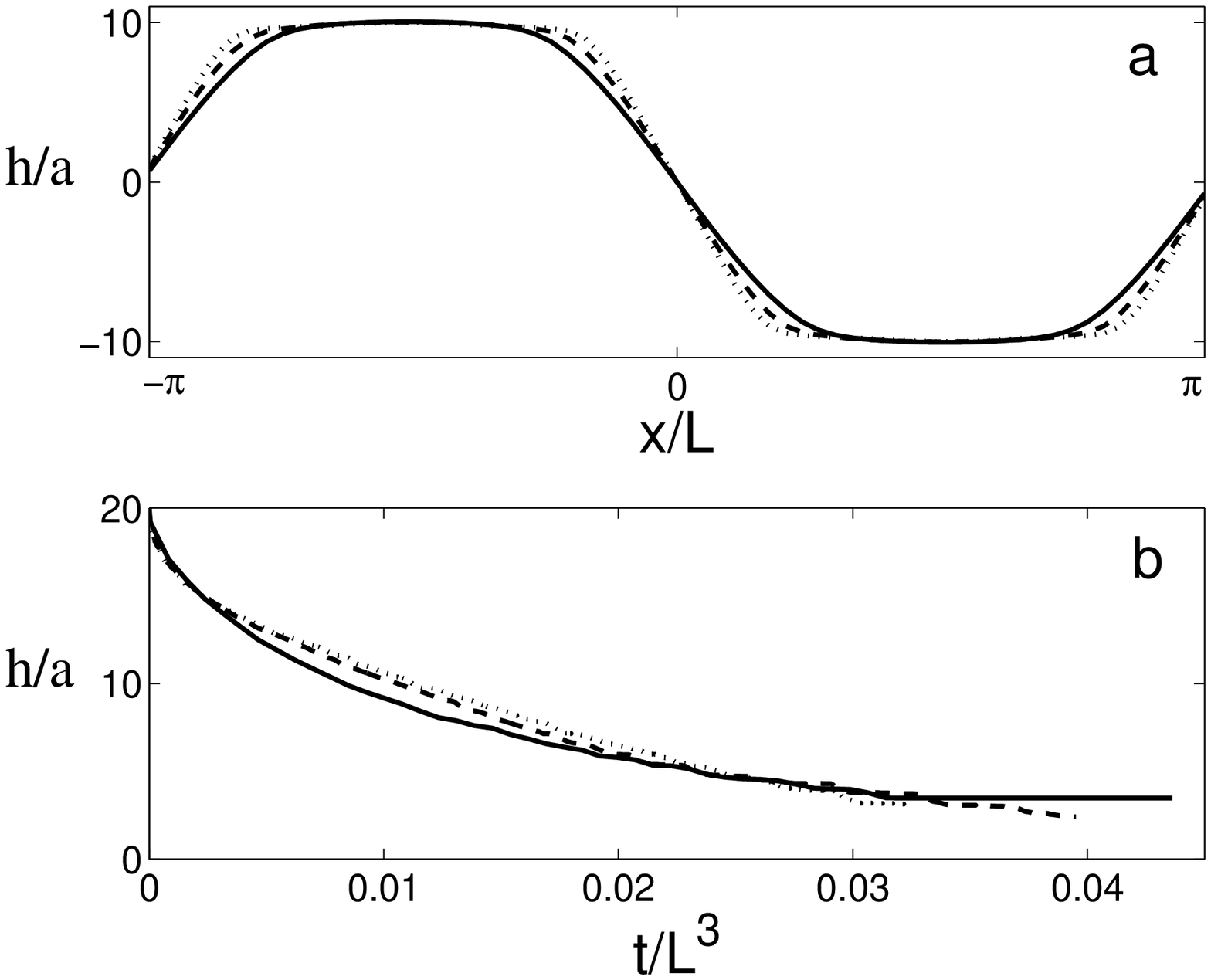}}
\vspace{1mm} \caption{Bidirectional sinusoidal profiles in the
case of stronger step-step interactions. Wave lengths shown are
$L=64$ (solid), $L=128$ (dashed) and $L=192$ (dotted). a) Cross
sections of the profiles with different wave lengths which have
relaxed to half of their initial amplitude. The cross sections are
along the $y=-L/4$ line (peak to valley). b) Amplitude decay of
the different wave lengths as a function of scaled time $t/L^3$.}
 \label{high_g_egg}
\end{figure}

Figure \ref{high_g_egg} shows results for a much stronger
repulsive interactions between steps, where the scaling law
(\ref{shape_scaling}) clearly does not hold. Fig.\
\ref{high_g_egg} (a) shows profiles of different wavelengths which
have relaxed to half of the initial amplitude. Here profiles with
a smaller wavelength have small facets and moderate slopes. This
happens because repulsion between steps becomes increasingly
important as the profile wavelength is reduced. At long
wavelengths the weak step-step interaction limit of Fig.\
\ref{low_g_egg} is recovered. Fig.\ \ref{high_g_egg} (b) shows the
different amplitudes as a function of scaled time $t/L^3$.

\section{Conclusions}
We proposed a novel continuum model for the evolution of sub
roughening crystal surfaces. Our model, which we term
Configurational Continuum, is derived directly from the underlying
dynamics of atomic steps. Unlike conventional continuum models,
Configurational Continuum is fully consistent with the step
dynamics and is capable of handling singular surface features such
as facets and corners.

The key idea in our model is to view the continuous surface
profile as the envelope of an ensemble of step configurations
which are all consistent with the continuous profile.  Knowing the
ensemble envelope, it is always possible to reconstruct individual
configurations and evolve them in time. This evolution of
individual configurations determines the evolution of the ensemble
envelope which within our model is interpreted as the evolution of
the profile. The continuum limit in our model is naturally
realized because the continuous profile induces a continuum of
possible step configurations.

Like other continuum models, Configurational Continuum has a
computational advantage over the underlying discrete step model.
When solved on a computer, it is possible to use a sparse grid in
regions where the profile is very smooth. However, near corners or
facets our model requires a fine discretization grid. The fine
grid is necessary in order to faithfully reconstruct the step
configurations which are possible near a singular point. Our model
thus has the important property of being capable of describing
step flow on different scales in a consistent way. In smooth
surface regions, Configurational Continuum provides a coarse
grained numerical description of surface dynamics. However it is
still capable of accounting for the unique behavior of steps near
singular points or lines.

The problem of connecting between different scales in dynamical
systems is not limited to the evolution of surfaces. This problem
is widespread in physics, engineering and biology as well as in
other fields. Our hope is that the ideas at the basis of
Configurational Continuum can be applied in other multi-scale
problems.

\appendix
\section{Failure of the cusp rounding method.}
\label{cusp_rounding_appendix} In this appendix we give an example
which shows how rounding of the surface free energy cusp can lead
to erroneous solutions for surface evolution. As an example system
we again choose the crystalline cone studied in Refs.\
\cite{cone_prl,cone}. For simplicity we consider the diffusion
limited case. This system is convenient since it exhibits scaling.
In the scaling limit conventional continuum modelling of the cone
becomes exact \cite{cone_prl,cone} and we can concentrate on
effects introduces by the cusp rounding method. We start by
solving a model with a rounded free energy cusp and later compare
the solution of this model with the relevant discrete step model
and with the solution of the Configurational Continuum.

In the conventional continuum approach, the continuous free energy
density of sub roughening surfaces has a cusp singularity at the
high symmetry orientation.  In a coordinate system where $(x,y)$
is the high symmetry plane and $h(x,y)$ is the surface profile,
the projected (on $(x,y)$) surface free energy density assumes the
form \cite{Gruber_JPCS28}:
\begin{equation}
{\mathcal F}\left(x,y\right)={\mathcal F}_0+\Gamma\left|\nabla h(x,y)\right|+\frac{G}{3}\left|\nabla h(x,y)\right|^3\;.
\label{cusped_free_energy}
\end{equation}
This form is the continuum analog of the free energy of an array
of steps with line tension $\Gamma$ and an inverse square
repulsive step-step interaction of strength $G$.

The singular nature of the surface free energy complicates the
modelling of surface evolution. Several authors have tried to
overcome this problem by rounding the cusp with a small parameter
$\alpha$:
\begin{equation}
{\mathcal F}_{\alpha}\left(x,y\right)={\mathcal
F}_0+\Gamma\left[\left(\nabla
h(x,y)\right)^2+\alpha^2\right]^{1/2}+\frac{G}{3}\left[\left(\nabla
h(x,y)\right)^2+\alpha^2\right]^{3/2}\;.
\label{rounded_free_energy}
\end{equation}
The hope behind this regularization scheme is that in the limit
$\alpha\rightarrow 0$ the resulting model captures the correct
surface dynamics.

Surface dynamics is derived from ${\mathcal F}_{\alpha}$ as
follows. Taking the functional derivative of ${\mathcal
F}_{\alpha}$ we obtain the surface chemical potential:
\begin{equation}
\mu_{\alpha}=\frac{\delta {\mathcal F_{\alpha}}}{\delta h}=-\Omega \left(\frac{\partial}{\partial x}\frac{\partial {\mathcal F}_{\alpha}}{\partial h_x}+\frac{\partial}{\partial y}\frac{\partial {\mathcal F}_{\alpha}}{\partial h_y} \right)\;,
\label{continuus_mu}
\end{equation}
where $h_x=\partial h/\partial x$ and $h_y=\partial h/\partial y$.
For a radially symmetric profile
$h(r,t)$ we find that
\begin{equation}
\mu_{\alpha}=-\frac{\Omega}{\sqrt{h_{r}^2+\alpha^2}}\left[\left(\frac{h_{r}}{r}+h_{rr}\right)\left(\Gamma+G\left(h_{r}^2+\alpha^2\right)\right)+h_{r}^2h_{rr}\left(G-\frac{\Gamma}{h_{r}^2+\alpha^2}\right)\right]\;,
\label{rounded_chemical_potential}
\end{equation}
where $h_{r}=\partial h/\partial r$ and $h_{rr}=\partial^2
h/\partial r^2$.

In diffusion limited kinetics variations in the chemical potential give rise to
currents which are proportional to the chemical potential
gradient. For our radially symmetric profile we can consider only
the radial component of this current
\begin{equation}
J=-\frac{D_s \tilde{C}^{eq}}{k_BT}\frac{\partial
\mu_{\alpha}}{\partial r}\;.
\end{equation}
The dynamic equation for the profile can now be written using the
continuity equation
\begin{equation}
\frac{\partial h}{\partial t}=-\Omega\cdot\nabla J=\frac{\Omega
D_s \tilde{C}^{eq}}{k_BT}\frac{1}{r}\frac{\partial}{\partial
r}\left(r\frac{\partial \mu_{\alpha}}{\partial r}\right)\;.
\label{round_cusp_dyn_eq}
\end{equation}

The $\alpha \rightarrow 0$ limit of Eq.\ (\ref{round_cusp_dyn_eq})
automatically gives the correct equilibrium crystal shapes because
Eq.\ (\ref{rounded_free_energy}) goes to Eq.\
(\ref{cusped_free_energy}) in this limit. We want to check whether
this limit also gives the correct (consistent with step flow)
dynamics.

\begin{figure}[h]
\centerline{ \epsfxsize=80mm
\epsffile{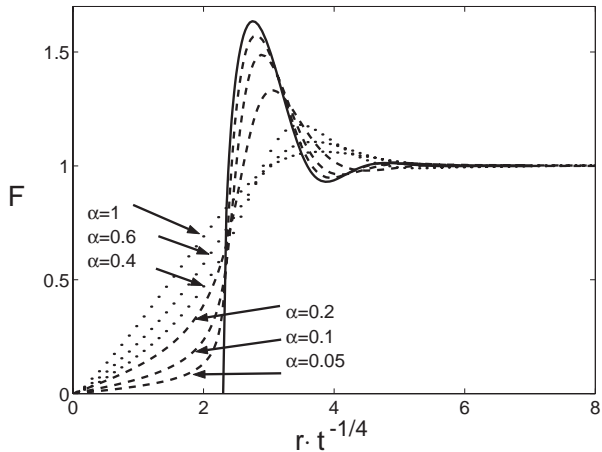}} \caption{Scaling
functions of the slope for different values of the cusp rounding
parameter. Dots show scaling functions obtained from a direct
solution of the dynamic equation (\ref{round_cusp_dyn_eq}). Dashed
lines show solutions of the scaling equation
(\ref{round_cusp_scaling_eq}). The solid line shows the estimated
$\alpha=0$ solution of Eq.\ (\ref{round_cusp_scaling_eq}).}
\label{round_cusp_scaling}
\end{figure}

We applied Eq.\ (\ref{round_cusp_dyn_eq}) to a crystalline cone
$h(r,t=0)\sim -r$. The similarity sign here indicates that the tip
of the initial profile was smoothed in order to have an analytic
surface. Analyticity at the origin was also used as a boundary
condition for the surface evolution. Numerical solutions show
that, at long times the profile slope obeys a scaling law
\begin{equation}
h_r(r,t)=-F(r t^{-1/4})\;. \label{cusp_rounding_scaling_ansatz}
\end{equation}
This behavior agrees with the scaling properties exhibited by a
discrete step flow model of the same surface structure
\cite{cone_prl,cone}. In Fig.\ \ref{round_cusp_scaling} we show
the resulting scaled slopes (dots) for different values of the
cusp rounding parameter $\alpha$. These long time solutions are
not sensitive to the initial smoothing of the cone. As $\alpha$ is
reduced we observe the appearance of a very flat region around the
origin. This flat region supposedly becomes a true facet in the
$\alpha=0$ limit.

For small values of $\alpha$, solutions of the dynamic equation
(\ref{round_cusp_dyn_eq}) approach scaling very slowly. For this
reason it becomes increasingly difficult to probe the $\alpha=0$
scaling state. In order to reach smaller values of $\alpha$ we
continued in the following way. We assumed that the scaling ansatz
(\ref{cusp_rounding_scaling_ansatz}) holds and used it to
transform Eq.\ (\ref{round_cusp_dyn_eq}) into an ordinary
differential equation for the scaling function $F(\xi)$ in the
scaling variable $\xi=rt^{-1/4}$. Replacing $h_r(r,t)$ in Eq.\
(\ref{round_cusp_dyn_eq}) by the scaling function $F(\xi)$ we
obtain the following equation:
\begin{eqnarray}
-\frac{1}{4}F'&=&\frac{\Omega D_s
\tilde{C}^{eq}}{k_BT}\frac{d}{d\xi}\left[\frac{1}{\xi}\frac{d}{d\xi}\left(\xi\frac{d\eta_{\alpha}}{d\xi}\right)\right]\;,
\nonumber \\
\eta_{\alpha}&=&-\frac{\Omega}{\sqrt{F^2+\alpha^2}}\left[\left(\frac{F}{\xi}+F'\right)\left(\Gamma
+G\left(F^2+\alpha^2\right)\right)\right. \nonumber \\
&&\;\;\;\;\;\;\;\;\;\;\;\;\;\;\;\;\;\;\;\;\left.+F^2F'\left(G-\frac{\Gamma}{F^2+\alpha^2}\right)\right]\;,
 \label{round_cusp_scaling_eq}
\end{eqnarray}
with $F'=dF/d\xi$.

Solutions of the above equation for the large values of $\alpha$
agree with the scaling states of the dynamic equation
(\ref{round_cusp_dyn_eq}). The dashed lines in Fig.\
\ref{round_cusp_scaling} show the resulting scaling functions for
smaller values of $\alpha$. Finally in order to determine the
$\alpha=0$ limit we estimated the scaled position of the facet
edge from the $\alpha\neq 0$ solutions. This position
selects\cite{cone_prl,cone} a unique scaling solution for the
$\alpha=0$ case of Eq.\ (\ref{round_cusp_scaling_eq}). The
resulting scaling function is shown by the solid line in Fig.\
\ref{round_cusp_scaling}. By our procedure this function is an
approximation for the true scaling function of the system
according to the cusp rounding method.

\begin{figure}[h]
\centerline{ \epsfxsize=80mm
\epsffile{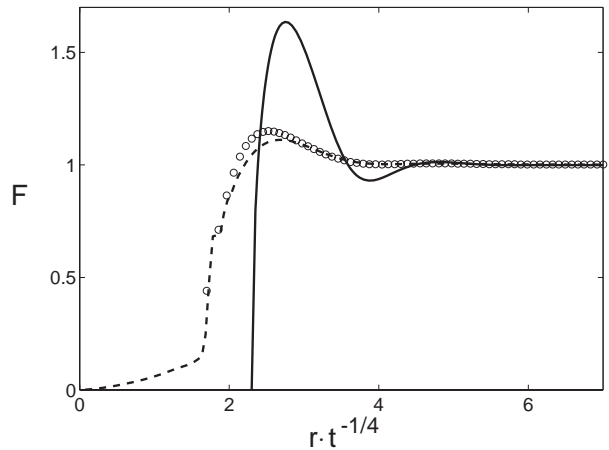}} \vspace{1mm}
\caption{Comparison between the scaling function predicted by the
cusp rounding procedure (solid), scaling function from a one
dimensional solution of the Configurational Continuum (dashed) and
the scaled density of steps in the discrete model (circles).}
\label{cusp_rounding_comparison}
\end{figure}

The solid line in Fig.\ \ref{round_cusp_scaling} should be
compared with the behavior of a system of discrete steps. For this
purpose we introduce the following step model:
\begin{eqnarray}
\frac{dr_1}{dt}&=&\frac{D_s C^{eq}_{\infty} \Omega}{k_B T
r_1}\frac{\mu_1-\mu_2}{\ln\left(r_1/r_2\right)}\;,
\nonumber \\
\frac{dr_n}{dt}&=&\frac{D_s C^{eq}_{\infty} \Omega}{k_B T
r_n}\left(\frac{\mu_n-\mu_{n+1}}{\ln\left(r_n/r_{n+1}\right)}-\frac{\mu_{n-1}-\mu_{n}}{\ln\left(r_{n-1}/r_n\right)}\right)\;,
\;\;\; n>1, \nonumber \\
\mu_n &=& \frac{\Omega \Gamma}{r_n} + \Omega G \left[
  \frac{2r_{n+1}}{r_{n+1}+r_n} \cdot
  \frac{1}{\left(r_{n+1}-r_n\right)^3} +\frac{r_{n+1}}{\left(r_{n+1}^2-r_n^2\right)^2} \right. \nonumber \\
  &&-\left.\frac{2r_{n-1}}{r_n+r_{n-1}} \cdot
  \frac{1}{\left(r_n-r_{n-1}\right)^3}+\frac{r_{n-1}}{\left(r_{n}^2-r_{n-1}^2\right)^2}
\right]\;.
\label{cone_step_velocity}
\end{eqnarray}
This is the same step model studied in section
\ref{config_cont_application} and in Refs. \cite{cone_prl,cone}
with modified step chemical potentials. The modification has a
small effect on the model behavior and does not introduce any
qualitative changes. In particular, this step model obeys the same
scaling properties that were studied in Refs.
\cite{cone_prl,cone}, i.e., the density of steps in this model
scales according to $D(r,t)=F_{discrete}(rt^{-1/4})$. In addition,
applying the scaling analysis of Refs.\ \cite{cone_prl,cone} to
this modified step model results in an ordinary differential
equation for the scaling function $F_{discrete}$ which is
identical to the $\alpha=0$ limit of Eq.\
(\ref{round_cusp_scaling_eq}). This fact gives us a basis for
comparison between the step model and the cusp rounding scheme.
Identifying the step density of the discrete model with the slope
of the continuum model we can finally compare the limiting
solution from the cusp rounding method with the scaled density of
steps. In Fig.\ \ref{cusp_rounding_comparison} we show that these
two functions do not coincide. The scaled position of the facet
edge in the cusp rounding method is about $40\%$ larger than the
one in the discrete system. This means that the difference between
the two facets in real space diverges at long times. The height at
the origin according to the cusp rounding method will suffer from
the same errors. By assuming analyticity of the profile throughout
the limiting procedure of the cusp rounding method, we have
imposed erroneous boundary conditions at the facet edge.

Figure \ref{cusp_rounding_comparison} also shows a one dimensional
solution of the Configurational Continuum model for this cone
system. The Configurational Continuum model predicts scaling of
the slope as well. The discrepancy between the resulting scaling
function and the discrete step system is much smaller and is
consistent with what one would expect from discretization errors.



\end{document}